\documentclass[showpacs,prl,aps,twocolumn]{revtex4}

\usepackage{bm}
\usepackage{amsmath}
\usepackage{amssymb}
\usepackage{latexsym}
\usepackage{amsfonts}
\usepackage{epsfig}
\usepackage{psfrag}
\usepackage{color}
\usepackage{dsfont}

\newcommand{\ket}[1]{\left|#1\right>}
\newcommand{\bra}[1]{\left<#1\right|}

\newcommand{\nn}{\nonumber\\}

\newcommand{\bea}{\begin{eqnarray}}
\newcommand{\ea}{\end{eqnarray}}
\newcommand{\eea}{\end{eqnarray}}

\definecolor{grey}{rgb}{0.5, 0.5, 0.5}
\definecolor{dgreen}{rgb}{0.0, 0.5, 0.0}
\definecolor{violet}{rgb}{0.5, 0.0, 0.5}
\definecolor{orange}{rgb}{1.0, 0.5, 0.0}

\begin{document}

\title{Charge Qubit Purification by an Electronic Feedback Loop}

\author{Gerold Kie{\ss}lich}
\email{gerold.kiesslich@tu-berlin.de}
\author{Gernot Schaller}
\author{Clive Emary}
\author{Tobias Brandes}

\affiliation{Institut f\"ur Theoretische Physik, Hardenbergstra{\ss}e 36,
Technische Universit\"at Berlin, D-10623 Berlin, Germany}


\begin{abstract}
We propose the manipulation of an isolated qubit by
a simple instantaneous closed-loop 
feedback scheme in which a time-dependent electronic 
detector current is directly back-coupled into qubit parameters.
As specific detector model we employ a capacitively coupled
single-electron transistor.
We demonstrate the stabilization of 
pure delocalized qubit states above a critical detector-qubit coupling. 
This electronic purification is independent of the initial qubit
state and is accomplished after few electron jumps through the detector. 
Our simple scheme can be used for the efficient and robust
initialization of solid-state qubits 
in quantum computational algorithms at arbitrary temperatures.   
\end{abstract}

\pacs{
03.65.Ta,  
05.40.-a,  
73.23.-b,  
85.35.-p   
}

\maketitle


The ability to measure and control individual quantum systems (e.g. qubits) is a major requirement for
contemporary and future information technologies \cite{WIS10}.
The measurement process -- the contact between classical
measurement device and nano-scale object -- plays a key role in controlling quantum
systems  and mostly introduces
a source of decoherence.
State-of-the-art measurement technologies in solid-state environments mainly rely on the
change of current flow through an on-chip detector due to
Coulomb repulsion with the charges in the measured system (see e.g. Refs.~\cite{GUS06}).  
The continuous monitoring of charge qubits by such charge detectors
yields relaxation and decoherence as studied  experimentally e.g. in Ref.~\cite{PET10}
and theoretically in Refs.~\cite{theory}.
However, the detector signal contains detailed information on the
qubit dynamics which can be used for its control.

An elegant control strategy, which has been
successfully applied in a wide variety of physical and engineering
systems \cite{SCH08a}, is stabilization of favored system states due to feedback.
For quantum systems such as qubits, several theoretical proposals to
enhance quantum coherence by feedback have been made.
These are mainly based upon sophisticated techniques, e.g.,	
the on-chip conversion of detector outputs into qubit density matrix
evolutions \cite{RUS02,WAN07a} or monitoring the phase of qubit
oscillations by detector current quadratures \cite{KOR05}.
However, a much simpler scheme was recently used in the first
experimental use of feedback to increase quantum coherence of a spin
qubit \cite{BLU10}.

\begin{figure}[t]
\includegraphics[width=.55\textwidth]{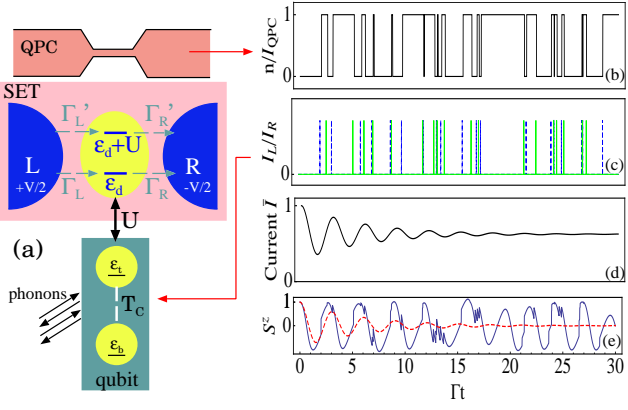}
\caption{(color online). {\bf (a)} Sketch of the device: charge qubit
and single-electron transistor (SET) as capacitively coupled detector; 
a quantum point contact (QPC) is attached to the SET to resolve $I_L(t)$ and $I_R(t)$ for the
closed-loop feedback. {\bf (b)} Single realization of the SET occupation $n(t)$ measured by the QPC current $I_{\textrm{QPC}}(t)$. 
{\bf (c)} Single realizations of SET currents $I_L(t)$ (blue, solid) and $I_R(t)$ (dashed, green).
{\bf (d)} SET current $\langle \bar I\rangle (t)$, ensemble averaged $\langle .\rangle$ 
and time averaged $\bar I=\frac{1}{\tau}\int_t^{t+\tau}I(t')dt'$ with $I(t)=\frac{1}{2}[I_L(t)+I_R(t)]$,
exhibits damped oscillations caused by the coupling to the qubit.
{\bf (e)} Bloch vector component $S^z(t)$ (qubit occupation): ensemble average (red, dashed) shows oscillations damped by the presence of the SET, 
single realization (blue, solid) exhibits free evolution interrupted whenever an electron jumps at the SET.
Parameters: $\Gamma'/\Gamma =$ 4, $T_c/\Gamma=$ 1, $\varepsilon =$ 0,
$U/\Gamma =$ 0.5, measurement time $\Gamma\tau =$ 0.01, initial state: $S^z(0)=$ 1,
$n(0)=$ 0.
}
\label{fig1}
\end{figure}

Here, we study a charge qubit consisting of a single electron 
confined in a two-state system which is continuously monitored by the
current through a single-electron transistor (SET) [see Fig.~\ref{fig1}].
The SET current will be directly fed back into qubit parameters which allow the stabilization of arbitrary pure qubit states.


\noindent
{\em Model.} -- The SET electronic level is shifted 
depending on where the electron is located in the two-level system due
to Coulomb interaction.
Moreover, the charge qubit electron is assumed to be coupled to thermal phonons of the surrounding matrix material.
The full Hamiltonian reads 
\bea
H&=&H_{\textrm{SET}}+H_{\textrm{q}}+\frac{1}{2}d^\dagger d\big[U\hat\sigma_z+(U_t+U_b)\mathds{1}\big],\\
H_{\textrm{q}}&=&\frac{1}{2}\varepsilon\hat\sigma_z+T_c\hat\sigma_x
+\sum_Q\bigg[\frac{1}{2}\hat\sigma_zg_QA_Q+\Omega_Qa_Q^\dagger a_Q\bigg],\nonumber
\eea
with $\varepsilon\equiv\varepsilon_t-\varepsilon_b$ and $A_Q\equiv a_{-Q}+a_Q^\dagger$. The SET
with Hamiltonian $H_{\textrm{SET}}=\varepsilon_dd^\dagger d+\sum_{ka
=L,R}[\varepsilon_{ka}c^\dagger_{ka}c_{ka}+(t_{ka}c_{ka}d^\dagger
+\textrm{h.c.})]$ contains
a single electronic level (with 
creation/annihilation operators $d^\dagger$/$d$) which is 
electrostatically coupled to the charge qubit levels by \mbox{$U\equiv U_t -
U_b \ge 0$}. 
The charge qubit Hamiltonian is given in the Bloch representation with Pauli
matrices 
\mbox{$\hat\sigma_z=\ket{t}\bra{t}-\ket{b}\bra{b}$} and 
$\hat\sigma_x=\ket{t}\bra{b}+\ket{b}\bra{t}$.
The phononic modes with wave vector $Q$ and energy $\Omega_Q$ are
diagonally coupled to the qubit assuming weak coupling $T_c$.


\noindent
{\em Equations of Motion.} -- The standard Born-Markov-secular approximation yields a quantum master equation
$\dot{\rho}(t)=\mathcal{L}\rho (t)$ for the system (SET + qubit)
density matrix $\rho (t)$
with the Liouvillian superoperator
$\mathcal{L}$ 
where the
SET leads and the phonon environment are treated in thermal
equilibrium. 
Due to the secular approximation the 4 populations and 2 coherences
contained in $\rho (t)$ decouple in the 
energy eigenbasis.
In the Markovian limit, tunneling rates and bath expectation values are
all evaluated at the transition frequencies of the system. 
Since the SET
Coulomb shift depends on the position of the charge qubit electron, this
introduces for finite SET bias effective tunneling rates $\Gamma_a$ and
$\Gamma_a'$ for lower and upper electron position, respectively, such
that the (time and ensemble averaged) SET current comprises damped 
oscillations for $\Gamma_a'\neq\Gamma_a$, see Fig.~\ref{fig1}(d).
These are caused by the qubit evolution [Fig.~\ref{fig1}(e)].
With the SET occupation \mbox{$n(t)=\textrm{Tr}[\hat{n}\rho (t)]$} and the
Bloch vector components conditioned on whether the SET is occupied, 
$S_1^\alpha (t)
\equiv\textrm{Tr}\big[(\hat{n}\otimes\hat{\sigma}_\alpha)\rho (t)\big]$ and
$S_0^\alpha (t)
\equiv\textrm{Tr}\big\{[(1-\hat{n})\otimes\hat{\sigma}_\alpha]\rho (t)\big\}$
($\alpha =x,y,z$) respectively, such
that $S^\alpha (t)= S_0^\alpha (t) + S_1^\alpha (t)$, the quantum master
equation has the following convenient vector form (omitting the time dependence):
\bea
\dot n&=&\gamma^+_L-\gamma^+ n-\gamma_L^-S_0^z+\gamma_R^-S_1^z,\nn
\dot{\mathbf{S}}_0&=&\mathbf{A}_0\times\mathbf{S}_0+\mathcal{D}_L\mathbf{S}_0+\mathcal{B}_R\mathbf{S}_1
+\mathbf{n}_0n+(\eta_r,0,\gamma_L^-)^T,\nn
\dot{\mathbf{S}}_1&=&\mathbf{A}_1\times\mathbf{S}_1+\mathcal{D}_R\mathbf{S}_1+\mathcal{B}_L\mathbf{S}_0
+\mathbf{n}_1n-(0,0,\gamma_L^-)^T,
\label{eq:ME_compact}
\eea
with  \mbox{$\mathbf{S}_i (t)\equiv (S_i^x(t),S_i^y(t),S_i^z(t))^T$}, \mbox{$\gamma_a^\pm\equiv\frac{1}{2}(\Gamma_a\pm\Gamma_a')$}, \mbox{$\gamma^\pm\equiv\gamma_L^\pm+\gamma_R^\pm$},  
\mbox{$\mathbf{n}_0=-\mathbf{n}_1\equiv-(\eta_r,0,\gamma^- )^T$}, and 
\bea
\mathcal{D}_a&\equiv&\left(
\begin{array}{ccc}
-\gamma_a^+-\eta_p & 0 & \eta_++\eta_-\\
0 & -\gamma_a^+-\eta_p & 0 \\
0 & 0 & -\gamma_a^+\\
\end{array}
\right),\nn
\mathcal{B}_a&\equiv&\textrm{Diag}\left(
\sqrt{\Gamma_a\Gamma_a'},\sqrt{\Gamma_a\Gamma_a'},\gamma^+_a\right),
\label{eq:coupling}
\eea
with the phonon rates $\eta_p$ and $\eta_r\equiv\eta_+-\eta_-$ explicitly defined in Ref.~\cite{KIE07b}.
For zero qubit bias ($\varepsilon =0$) they are $\eta_r=g\pi
T_ce^{-2T_c/\omega_c}$ and $\eta_p =\eta_r\coth{(\beta T_c)}$ with the
dimensionless coupling strength $g$, the cut-off
frequency $\omega_c$, and the inverse temperature $\beta$.
The free dynamics of ${\mathbf{S}}_j(t)$ in
Eqs.~(\ref{eq:ME_compact}) can be considered as a precession around vector
\mbox{$\mathbf{A}_j\equiv -\big(2T_c,0,\varepsilon +jU\big)^T$} with frequency 
\mbox{$\Delta\equiv\sqrt{\varepsilon^2+4T_c^2}$} ($j=0$) or
\mbox{$\Delta'\equiv\sqrt{(\varepsilon+U)^2+4T_c^2}$} ($j=1$), respectively.
The effect of dissipation is contained in $\mathcal{D}_a$: damping of
all Bloch vector components by the 
presence of the SET with 
$\gamma^+_L$ or $\gamma^+_R$ and by the electron-phonon coupling which
leads to an additional decay and scattering of the $S^{x/y}_i(t)$.
The conditioned Bloch vectors ${\mathbf{S}}_j(t)$ are mutually coupled
via the diagonal matrices $\mathcal{B}_{L/R}$, i.e., by electron hopping onto the SET and out of it, respectively.
We provide an analytical
solution in the online supplement \cite{sup}.
The SET occupation couples to the qubit and is not
affected by its dynamics for $\Gamma_a=\Gamma_a'$.
We note that the Coulomb interaction $U$ enters Eqs.~(\ref{eq:ME_compact}) implicitly in $\Gamma'$ and explicitly in $\mathbf{A}_1$. 
Whenever we set $U=$ 0 in the following we do not exclude $\Gamma'\neq\Gamma$ in order to maintain the coupling but rather neglect the physical 
back-action on the qubit.


\begin{figure}[t]
\includegraphics[width=.48\textwidth]{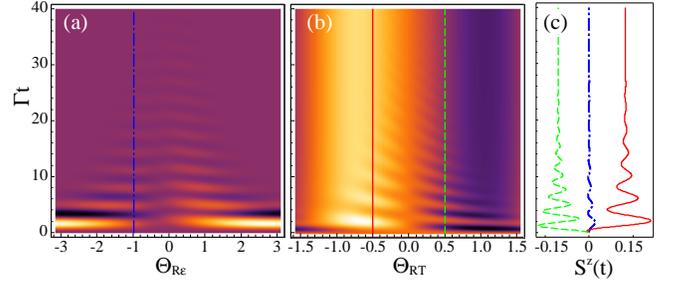}
\caption{(color online). Bloch vector component $S^z(t)$ vs. time $\Gamma t$ and feedback parameter
$\Theta_{R\varepsilon}$ {\bf (a)} and $\Theta_{RT}$ {\bf (b)};
{\bf (c)} $S^z(t)$ cross-sections at $\Theta_{R\varepsilon}=-$1 (blue,
dash-dotted),
$\Theta_{RT}=-$0.5 (red, full), 0.5 (green, dashed).
Parameters: $\varepsilon =$ 0, $U/\Gamma =$ 0.5, $\Gamma'/\Gamma =$ 3,
$T_c/\Gamma =$ 1. Initial state is the complete statistical mixture with empty
SET. With feedback damped qubit oscillations can be
induced which would not occur for vanishing SET-qubit
coupling $\Gamma =\Gamma'$. 
}
\label{fig2}
\end{figure}


\noindent
{\em Detector counting statistics and active measurement.} --
The information about the number of charges $n_a$ transferred  through a
multi-lead conductor (our SET detector) at the $a$-th lead   can be gained by the help
of $n_a$-resolved quantum master equations (see e.g. Ref.~\cite{SCH09c}
for detailed discussions).
The unraveled Fourier-transformed master equation for unidirectional
transport is 
$\dot\rho =[\mathcal{L}_0+\sum_a e^{i\chi_a}\mathcal{J}_a]\rho \equiv\mathcal{L}(\{\chi_a\})\rho$,
where the $\mathcal{J}_a$($\mathcal{L}_0$) denote jump (non-jump) operators  and
$\chi_a$ are the counting fields
and the system state $\rho$ is not actively controlled ({\em passive} measurement).
In contrast, when active control operations on the system
depending on the measurement outcomes are performed after single jump events, the 
Fourier-transformed master equation generalizes to
\bea
\frac{d}{dt}\rho\big(\{\chi_a\},t\big)
=\bigg[\mathcal{L}_0+\sum_ae^{i\chi_a}e^{\hat{\mathcal{K}}_a}\mathcal{J}_a\bigg]\rho \big(\{\chi_a\},t\big)
\label{eq:master}
\eea
where the $\hat{\mathcal{K}}_a$ describe the control operation \cite{WIS10}.
In contrast to passive measurements, in an active (control)
measurement the (scalar) counting fields $\chi_a$ are thus upgraded to
superoperators, $i\chi_a\to i\chi_a+\hat{\mathcal{K}}_a$.


\noindent
{\em Feedback scheme.} --
We choose the feedback of the SET current acting on $T_c$ or/and on
$\varepsilon$
which are accessible in present-days setups by appropriate gate voltages.
The corresponding feedback Hamiltonian is defined by
\mbox{$H_{\textrm{fb}}=\sum_{a=L/R}I_a(t)h_a$} with
$h_a\equiv \Theta_{a\varepsilon}\hat\sigma_z+\Theta_{aT}\hat\sigma_x$
and the feedback parameters $\Theta_{aT}$ and $\Theta_{a\varepsilon}$.
Experimentally, single jump events $I_{L/R}(t)$ may neither be
resolved nor measured independently in the SET circuit.
Therefore, we propose an additional quantum-point contact (QPC)
weakly attached to the SET as sketched in Fig.~\ref{fig1}(a).
Its telegraph-like current signal $I_{\textrm{QPC}}(t)$ provides
the occupation of the SET [Fig.~\ref{fig1}(b)].
For $\Gamma_a'>\Gamma_a$ the upwards/downwards steps are related to
single jumps into/out of the SET
and, consequently, $I_L(t)$/$I_R(t)$ are trains of $\delta$-kicks [Fig.~\ref{fig1}(c)].
The superoperators $\hat{\mathcal{K}}_{L/R}$ in  the feedback master
equation (\ref{eq:master}) are
then given by \mbox{$\hat{\mathcal{K}}_a\rho=-i\big[h_a,\rho\big]$}.
With feedback into $\varepsilon$/$T_c$, the diagonal matrices
$\mathcal{B}_{L/R}$ (\ref{eq:coupling}) become
rotation matrices around the qubit $x$/$z$-axis \cite{sup}, so that
whenever an electron jumps into/out of the SET the qubit state is
rotated by the corresponding angle.


\noindent
{\em Feedback-induced qubit oscillations.} --
As a remarkable effect of this Markovian feedback, we find
damped qubit oscillations even when 
the qubit is initially prepared in a completely mixed state (i.e., in the origin of the Bloch
sphere: $\mathbf{S}=\mathbf{0}$).
For low electron-phonon coupling without feedback \cite{sup} or
for vanishing SET-qubit coupling ($\Gamma'=\Gamma$, i.e., the 
qubit kicked by a random sequence of $\delta$-pulses)
the qubit would remain completely mixed.
Fig.~\ref{fig2} shows the oscillations in $S^z(t)$ for a feedback
from $I_R(t)$ [$I_L(t)$ provides similar results] 
into (a) $\varepsilon$ and (b) $T_c$.
The feedback into $\varepsilon$ [Fig.~\ref{fig2}(a)] induces
oscillations  
with the frequency
$\omega_Q=\frac{1}{4}\sqrt{(8T_c)^2-\gamma_S^2}$ ($\varepsilon =0$)
for $8T_c>\gamma_S$ where 
$\gamma_S=\frac{1}{2}(\sqrt{\Gamma}-\sqrt{\Gamma'})^2$ is the damping  due to SET-qubit coupling \cite{sup} and
additional feedback damping which increases with $\Theta_{R\varepsilon}$.
At large times, the qubit state is again completely mixed.
In contrast, for the  $T_c-$feedback [Fig.~\ref{fig2}(b)] a frequency
shift towards higher(lower) 
frequencies for $\Theta_{RT}>(<)$ 0 occurs,
and in the long-term limit a localization [$S^z(t\to\infty )\neq 0$] takes place.
The transient oscillations of the qubit occupation
[Fig.~\ref{fig2}(c)] can be directly measured as damped current
oscillations [compare Figs.~\ref{fig1}(d) and (e)].


\begin{figure}[t]
\includegraphics[width=.45\textwidth]{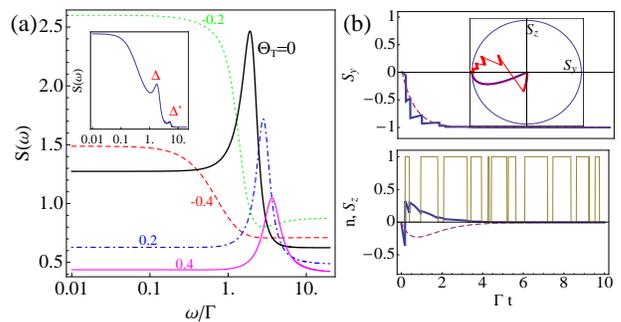}
\caption{(color online). {\bf (a)} SET noise spectrum $S(\omega )$
vs. $\omega$ for various feedback parameter
$\Theta_{T}=\Theta_{LT}=\Theta_{RT}=$ $-$0.4 (dashed, red), $-0.2$
(dotted, green), 0 (thick solid, black), 0.2 (dash-dotted, blue), 0.4
(thin solid, violet); {\bf (Inset)} $S(\omega )$ for $\Theta_{T}=$ 0, $\Gamma'/\Gamma
=$ 1.5, $U/\Gamma =$ 5. 
{\bf (b)} Single realization of Bloch vector components $S^{y,z}(t)$ and SET occupation $n(t)$
vs. time $\Gamma t$ with initially 
mixed state $\mathbf{S}=$ 0;
ensemble averaged quantities (dashed lines); {\bf (Inset)} Single realizations in the Bloch
sphere; measurement time $\Gamma\tau =$ 0.1, feedback parameter: $\Theta_T=-$0.45. Parameters: 
 $\varepsilon =$ 0, $U/\Gamma =$ 1, $\Gamma'/\Gamma =$ 9, $T_c/\Gamma =$ 1. 
}
\label{fig3}
\end{figure}


\noindent
{\em Noise spectrum.} -- The total noise spectrum of
the SET current is computed by (Ramo-Shockley theorem) \mbox{$S(\omega
)=\frac{1}{4}\sum_{a,b=L,R}S_{ab}(\omega )$} for a symmetric SET with \mbox{$S_{ab}(\omega
)=\textrm{Tr}\big\{[\mathcal{J}_{ab}+\mathcal{J}_a\Omega(\omega
)\mathcal{J}_b+\mathcal{J}_a\Omega(-\omega
)\mathcal{J}_b]\bar\rho\big\}$}
where $\Omega(\omega )=[-i\omega\mathds{1}-\mathcal{L}(0,0)]^{-1}$,
steady state $\bar\rho$ with $\mathcal{L}(0,0)\bar\rho =0$, and the jump operators
$\mathcal{J}_a\equiv\partial_{i\chi_a}\mathcal{L}(\chi_L,\chi_R)\vert_{(0,0)}$,
$\mathcal{J}_{a,b}\equiv\partial_{i\chi_a}\partial_{i\chi_b}\mathcal{L}(\chi_L,\chi_R)\vert_{(0,0)}$.
The spectrum provides detailed information on the qubit dynamics such as
oscillation frequencies and relaxation rates.
Without feedback,  $S(\omega )$ shows peaks at the bare qubit oscillation frequency $\Delta$ (for $\gamma_S\ll T_c$) and 
at the shifted frequency due to Coulomb interaction $\Delta'$ for
nonvanishing and weak SET-qubit coupling $\Gamma'/\Gamma\neq$ 1, see inset of Fig.~\ref{fig3}(a).
We can further use it to monitor the effect of feedback on the qubit
dynamics, as shown in Fig.~\ref{fig3}(a).
The most interesting feedback scenario is when both $I_L(t)$ and
$I_R(t)$ are simultaneously coupled back into $T_c$.
This yields a frequency shift linear in
$\Theta_T\equiv\Theta_{LT}=\Theta_{RT}$ 
[compare Fig.~\ref{fig2}(b)] of the $\Delta$-peak which is clearly
revealed in $S(\omega )$ [Fig.~\ref{fig3}(a)].
Moreover, at low frequencies the noise exhibits a pronounced
local minimum w.r.t. $\Theta_T$ at small negative values (not shown) which turns out
to correspond to the stabilization of pure qubit states.
%

\begin{figure}[t]
\includegraphics[width=.48\textwidth]{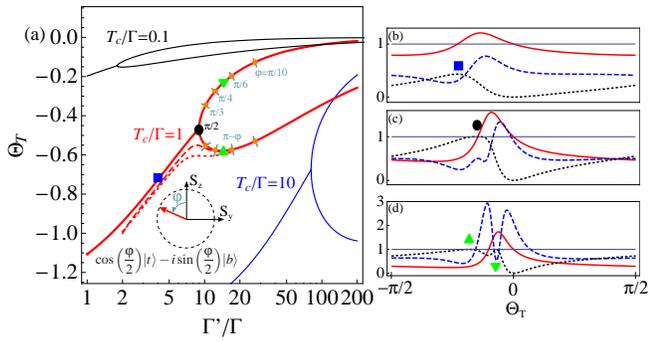}
\caption{(color online). {\bf (a)} Location of purity $P$ maxima in feedback parameter
plane ($\Theta_T$, $\Gamma'/\Gamma$), Eq.~(\ref{eq:pure_angle}).
Above a critical coupling  $\Gamma'/\Gamma \ge 1+8T_c/\Gamma$ (bifurcation) the qubit state is pure with $P=1$, 
otherwise $P<1$. 
At the bifurcation the qubit state is maximally delocalized with $S^y =-1$ ($\varphi =\pi /2$) and  above $S^z\neq
0$ with $0<\varphi <\pi$, $\varphi\neq\pi /2$ holds.  The bifurcation becomes blurred and the complete
delocalization is diminished
for $U/\Gamma =$ 0.5 (dashed line) and/or $\eta_r/\Gamma =$ 0.1,
$\beta T_c
=$ 1 (dotted line).
{\bf (b)-(d)} Normalized steady-state Current $I/I_0$ with
$I_0\equiv\frac{\Gamma +\Gamma'}{8}$ (red, solid curves), 
Fano factor $F=S(0)/(2I)$
(blue, dashed curves), purity $P=\textrm{Tr}(\rho^2)$ (black, dotted curves) vs. 
$\Theta_T$. Parameters: $\varepsilon =U=$ 0, $T_c/\Gamma =$ 1, $\Gamma'/\Gamma =$ 4 (b), 9 (c), 15 (d).
}
\label{fig4}
\end{figure}

\noindent
{\em Generation of pure states} --
The simulation of the quantum trajectory in Fig.~\ref{fig3}(b) indeed
confirms that by starting in a complete statistical mixture, the qubit will
finally approach the  pure state $\frac{1}{\sqrt{2}}(\ket{t}-i\ket{b})$ ($S^{y}=-1$) after a
few electron jumps through the SET.
In the following we discuss in more detail the stabilization of such pure qubit
states.
The steady state yields
\bea
\bar{n}&=&\frac{1}{2},\, \bar{S}^y=\gamma^-\frac{4c T_c}{4 T_c (4 T_c + sG)+c \gamma^+ (G+2\eta_p) },
\label{eq:steadyFB}
\eea
with $s\equiv\sin{(2\Theta_T)}$, $c\equiv1-\cos{(2\Theta_T)}$, and
\mbox{$G\equiv \frac{1}{2}(\sqrt{\Gamma}+\sqrt{\Gamma'})^2$}. The
$\bar{S}^x$-component is solely governed by 
electron-phonon relaxation \cite{sup} 
(we omit the $\bar{S}^z$-component for brevity).
We note that the SET occupation is not affected by the feedback due to
the high SET bias limit.
The steady-state purity [$P\equiv\textrm{Tr}(\bar\rho^2)=\sum_\alpha(\bar{S}^\alpha)^2$], i.e., the length of the Bloch vector 
for small electron-phonon coupling $\eta_p\ll\Gamma$ reads
\bea
P=(\gamma^-)^2c\frac{c G^2 + 
 8 T_c (4 T_c+sG)}{c \gamma^+ G + 
  4 T_c (4 T_c + s G)^2}.
\label{eq:pure}
\eea
First, for zero feedback ($\Theta_T=$ 0, $c=$ 0) the purity vanishes as
expected.
Second, above a critical coupling  $\Gamma'/\Gamma \ge 1+8T_c/\Gamma$
following from the condition  $B\equiv\sqrt{(\gamma^-)^2 - 16
T_c^2}\ge$ 0, two maxima of $P$ exist:
\bea
\cos^2{\big(\Theta_T^\pm\big)}=\frac{G[(\gamma^+)^2 + D + 12T_c^2] - 2 \gamma^+ D\pm 
  8 T_c^2 B }{
 G (\Gamma\Gamma' + 16T_c^2)},
\label{eq:pure_angle}
\eea
with $D\equiv G^2 + 4 T_c^2$ where the stationary qubit state has purity $P=1$.
Below the bifurcation ($B^2<$ 0) the state is partially mixed ($P <$ 1) and
\mbox{$\cos{(\Theta_T)}=-G(G^2+16T_c^2)^{-1/2}$} holds.
Fig.~\ref{fig4}(a) shows this bifurcation behavior of the maximal
purity for various $T_c$. 
The bifurcation point ($B=0$) corresponds to the
completely delocalized state with $S^y=-1$ ($\varphi =\pi /2$)
The sign of $S^y$ is determined by $\gamma^-$, compare Eq.~(\ref{eq:steadyFB}).
In the upper(lower) branch the pure qubit state becomes
increasingly localized with $S^z>$ 0 ($S^z<$ 0).
The effect of finite $U$ and the phononic environment onto the
steady-state purity is essentially a blurring of the bifurcation in
Fig.~\ref{fig4}(a) (dashed and dotted curves).
A complete delocalization ($S^y=-$1) as for the noninteracting case
hence cannot be obtained here anymore.
Nevertheless, asymptotic purification with increasing localization can be realized in the limit of large
SET-qubit coupling.


\noindent
{\em Steady-state SET transport.} --  Without feedback we obtain 
the steady-state current 
$I_0/e=(\Gamma +\Gamma')/8$ 
and the Fano factor 
$F\equiv\frac{S(0)}{2I}=\frac{1}{2}+2\frac{(\gamma_S+2\eta_p)(\Gamma -\Gamma')^2}{(8T_c)^2(\Gamma +\Gamma')}$.
The SET current is not affected by the phonons, whereas the noise
carries information about the interaction of the qubit with the SET
detector $\gamma_S$ and the phononic environment $\eta_p$.
The Fano factor is one-half for vanishing SET-qubit coupling ($\Gamma
=\Gamma'$) and diverges for large SET-qubit coupling ($\Gamma' /\Gamma\gg 1$) which results from the bistable
current behavior (quantum Zeno effect).
It also diverges for $T_c\to 0$ since the two subspaces decouple in
this limit.
Figs.~\ref{fig4}(b)-(d) present the $\Theta_T$-dependencies of the
steady-state current and the Fano factor at various detector-qubit couplings $\Gamma'/\Gamma$.
The current, in principle, follows the electron localization in the qubit.
At the critical coupling [Fig.~\ref{fig4}(c)] when $P=1$ the qubit state is maximally delocalized so that the
current approaches its no-feedback value $I_0$ and the
Fano factor displays a local maximum.
For larger couplings [Fig.~\ref{fig4}(d)] the purity maximum in the upper branch is accompanied with a current maximum and a Fano factor minimum.
Hence, the degree of the
qubit state purity can be very accurately monitored by varying the
feedback parameter $\Theta_T$.


\noindent
{\em Conclusions.} -- We have presented a simple way to
stabilize exact pure states and generate  charge oscillations in  an
individual charge qubit by applying an electronic feedback loop.
Even for moderate detector back-action and coupling to a dissipative environment
(e.g. phonons) an asymptotic purification can be achieved.
We note that in contrast to ground state relaxation just by cooling and strong electron-phonon coupling our method also works at high temperatures 
and, in principle, arbitrary 
pure states can be produced.
The detector current and its noise spectrum are shown to be sensitive monitors
for the effect of feedback.


Financial support by the DFG (\mbox{BR 1528/5}, \mbox{BR 1528/8}, \mbox{SCHA 1646/2-1}, and SFB 910) and 
helpful discussions with F. Haupt, K. Mosshammer and C. P{\"o}ltl 
are gratefully acknowledged.

\vspace{-6mm}

\end{document}


\title{Supplementary Material:\\ {\em Charge Qubit Purification by an
Electronic Feedback Loop}}

\author{G. Kie{\ss}lich}
\email{gerold.kiesslich@tu-berlin.de}
\author{G. Schaller}
\author{C. Emary}
\author{T. Brandes}

\affiliation{Institut f\"ur Theoretische Physik, Hardenbergstra{\ss}e 36,
Technische Universit\"at Berlin, D-10623 Berlin, Germany}

\maketitle


Equation numbers in the main text are referred to by an asterisk.


\section{Qubit Dynamics without feedback}

We rewrite Eqns.~(2*)  as vector equation $\dot{\mathbf{x}}=\mathcal{C}\cdot\mathbf{x}+\mathbf{v}$ with
\bea
\mathcal{C}&\equiv&\left(
\begin{array}{ccc}
-\gamma^+ & -\gamma_L^-\,\mathbf{e}_z^T & \gamma^-_R\,\mathbf{e}_z^T\\
\mathbf{n}_0 & \mathcal{A}_0+\mathcal{D}_L & \mathcal{B}_R\\
\mathbf{n}_1 & \mathcal{B}_L & \mathcal{A}_1+\mathcal{D}_R\\
\end{array}
\right),\\
\mathbf{x}&\equiv& (n,\mathbf{S}_0,\mathbf{S}_1)^T,\quad\mathbf{v}\equiv (\gamma^+_L,\gamma^+_r\mathbf{e}_x+\gamma^-_L\mathbf{e}_z,-\gamma^-_L\mathbf{e}_z)^T.\nonumber
\eea
For its solution we use the Laplace transform \mbox{$\tilde{\mathbf{x}}(z)=\big[z\mathds{1}-\mathcal{C}\big]^{-1}\cdot\big(\mathbf{x}_0+\mathbf{v}/z\big)$}
with the initial state \mbox{$\mathbf{x}_0\equiv (n_0,S^{x0}_0,S^{y0}_0,S^{z0}_0,S^{x0}_1,S^{y0}_1,S^{z0}_1)^T$}.

For symmetric coupling 
($\Gamma /2\equiv\Gamma_L=\Gamma_R$, $\Gamma' /2\equiv\Gamma_L'=\Gamma_R'$) and $\varepsilon =U=0$ we find

\bea
\tilde{n}(z)&=&\frac{A_1(z,n_0)+zA_2(z)+(\gamma^+ +2n_0z)A_3(z)}{2z[A_1(z,1)
+(\gamma^+ +z)A_3(z)]},\nn
\tilde{S}^x(z)&=&\frac{2}{g(z)}S^{x0},\quad
\tilde{S}^y(z)=2\frac{2T_c S^{z0} +zS^{y0}}{8 T_c^2 + z g(z)},\nn
\tilde{S}^z(z)&=&\frac{-4 T_c S^{y0} +  g(z)S^{z0}}{8 T_c^2 + z g(z)},
\label{eq:laplace}
\eea

with
\mbox{$A_1(z,n_0)\equiv\sqrt{\Gamma\Gamma'}[\Gamma\Gamma'+2n_0z(\gamma^+
+z)]-(\gamma^-)^2(\gamma^+ +2z)$}, \mbox{$A_2(z)\equiv 
\sqrt{\Gamma\Gamma'}\gamma^+ -\gamma^- [g(z)S^{z-}-4T_cS^{y-}]$}, 
\mbox{$A_3(z)\equiv 8T_c^2+(\gamma^+ +z)(\gamma^+ +2z)$},
$S^{y/z-}\equiv S_0^{y/z0}-S_1^{y/z0}$, 
and \mbox{$g(z)\equiv \gamma^+ -\sqrt{\Gamma\Gamma'} + 2 z$}.

The inverse Laplace transform  for $\varepsilon =U=$0 yields
\bea
S^x(t)&=&\frac{2\eta_r\big(1-e^{-\frac{1}{2}\gamma_Dt}\big)}{\gamma_S+2\eta_r\coth{(\beta
T_c)}}
+e^{-\frac{1}{2}\gamma_Dt}S^{x0},\nn
S^{y/z}(t)&=&e^{-\frac{1}{4}\gamma_Dt}\bigg[S^{y/z0}\cosh{(\omega_Qt)}
\mp\frac{\gamma_DS^{y/z0}-8T_cS^{z/y0}}{4\omega_Q}\sinh{(\omega_Qt)}\bigg],
\label{eq:bloch}
\eea
with $\omega_Q\equiv\frac{1}{4}\sqrt{\gamma_D^2-(8T_c)^2}$ (for
$\gamma_D\ll 8T_c$ damped qubit oscillations occur.) and 
$\gamma_D\equiv\gamma_S+2\eta_p$ where $\gamma_S\equiv\frac{1}{2}\big(\sqrt{\Gamma}-\sqrt{\Gamma'}\big)^2$
is the decoherence rate due to the SET-qubit coupling which was also found for a qubit coupled to a QPC in
Ref.~\cite{GUR97} where the respective 
rates are given by the QPC transmissions $T$ and $T'$.
For $U>0$ the expressions become rather cumbersome such that the decay can be non-exponential (see e.g. Ref.~\cite{EMA08} and discussions therein).
The steady state can be readily obtained by $\lim_{z\to
0^+}z\,\tilde{\mathbf{x}}(z)$ and yields (also for $\varepsilon ,U\neq$0)
\bea
\bar{n}=\frac{1}{2},\quad\bar{\mathbf{S}}=\bigg(\frac{2\eta_r}{\gamma_S+2\eta_r\coth{(\beta T_c)}},0,0\bigg)^T,
\label{eq:sx_phon}
\eea
with $\eta_r=g\pi T_ce^{-2T_c/\omega_c}$ being the phonon relaxation rate where $g$ is the dimensionless electron-phonon coupling strength and $\omega_c$ is a cut-off frequency \cite{BRA02}.
For zero phonon temperature ($\beta\to\infty$) the $x-$component becomes
$y/(1+y)$ ($y\equiv 2\eta_r/\gamma_S$) also discussed in Ref.~\cite{GUR03}; 
the qubit relaxes into a pure state for $y\gg 1$ (qubit ground state).
For large temperatures ($\beta\to 0$) the qubit will behave as statistical mixture.


\section{Feedback-modified equations of motion}

The feedback scheme modifies Eqns.~(2*) such
that $\mathcal{B}_a\to\mathcal{B}_a^f$,
$\mathbf{n}_0\to\mathbf{n}_0^f$, $\mathbf{n}_1\to\mathbf{n}_1^f$, and $\mathbf{v}\to\mathbf{v}^f$.
For $\Theta_{aT}=0$ we find $n_0^f=n_0$, $n_1^f=n_1$, $\mathbf{v}^f=\mathbf{v}$, and
\bea
\mathcal{B}_a^f&=&\left(
\begin{array}{ccc}
\cos{(
\Theta_{a\varepsilon} )}\sqrt{\Gamma_a\Gamma_a'} & \sin{(\Theta_{a\varepsilon} )}\sqrt{\Gamma_a\Gamma_a'} & 0\\
-\sin{(\Theta_{a\varepsilon} )}\sqrt{\Gamma_a\Gamma_a'} & \cos{(
\Theta_{a\varepsilon} )}\sqrt{\Gamma_a\Gamma_a'} & 0\\
0 & 0 & \gamma^+_R
\end{array}
\right),
\eea
which represents a rotation of the Bloch vector around the $z-$axis with angle $\Theta_{a\varepsilon}$. Respectively for $\Theta_{a\varepsilon}=0$ 
the matrix $\mathcal{B}_a^f$ rotates the Bloch vector  around the $x-$axis with angle $\Theta_{aT}$:
\bea
\mathcal{B}_a^f&=&\left(
\begin{array}{ccc}
\sqrt{\Gamma_a\Gamma_a'} & 0 & 0\\
0 & \cos{( 2\Theta_{aT})}\sqrt{\Gamma_a\Gamma_a'} &  \sin{( 2\Theta_{aT})}\gamma^+_a\\
0 & -\sin{( 2\Theta_{aT})}\sqrt{\Gamma_a\Gamma_a'} & \cos{(
2\Theta_{aT})}\gamma^+_a
\end{array}
\right),\nn
n_0^f&=&-\gamma^-_R\bigg(0,\sin{( 2\Theta_{RT})},\cos{(
2\Theta_{RT})}+\frac{\gamma^-_L}{\gamma^-_R}\bigg)^T,\nn
n_1^f&=&\gamma^-_L\bigg(0,\sin{( 2\Theta_{LT})},\cos{(
2\Theta_{LT})}+\frac{\gamma^-_R}{\gamma^-_L}\bigg)^T,\nn
\mathbf{v}^f&=&\bigg(\gamma^+_L,\gamma^-_L\mathbf{e}_z,-\gamma^-_L[\sin{(2\Theta_{LT})}\mathbf{e}_y+\cos{(2\Theta_{LT})}\mathbf{e}_z]\bigg)^T.
\eea

For symmetric coupling and $\varepsilon =U=0$ the $x-$components
decouple, and we can truncate the dynamical system to 5 equations:
\bea
\mathcal{C}^f=\left(\begin{array}{c|cc|cc}
-\gamma^+ & 0 & -\frac{\gamma^-}{2} & 0 & \frac{\gamma^-}{2}\\
\hline
-\frac{\gamma^-}{2}s_R & -\frac{\gamma^+}{2}-\eta_p & 2T_c & \frac{\sqrt{\Gamma\Gamma'}}{2}c_R & \frac{\gamma^+
}{2}s_R\\
-\frac{\gamma^-}{2}(1+c_R) & -2T_c & -\frac{\gamma^+}{2} &
-\frac{\sqrt{\Gamma\Gamma'}}{2}s_R & \frac{\gamma^+}{2}c_R\\

\hline
\frac{\gamma^-}{2}s_L & \frac{\sqrt{\Gamma\Gamma'}}{2}c_L & \frac{\gamma^+}{2}s_L & -\frac{\gamma^+}{2}-\eta_p & 2T_c\\
\frac{\gamma^-}{2}(1+c_L) & -\frac{\sqrt{\Gamma\Gamma'}}{2}s_L & \frac{\gamma^+}{2}c_L & -2T_c & -\frac{\gamma^+}{2}
\end{array}
\right),
\eea
with $s_a\equiv\sin{( 2\Theta_{aT})}$ and $c_a\equiv\cos{( 2\Theta_{aT})}$. 
